\documentclass[runningheads]{llncs}
\usepackage{graphicx}
\begin{document}
\title{V-Dream: Immersive Exploration of Generative Design Solution Space}
%
%
\author{Mohammad Keshavarzi\inst{1}
\and
Ardavan Bidgoli\inst{2}
\and
Hans Kellner\inst{3}
}
%
%
\institute{University of California, Berkeley \email{mkeshavarzi@berkeley.edu} \and
Carnegie Mellon University
\email{abidgoli@andrew.cmu.edu}
\and
Autodesk Research
\email{hans.kellner@autodesk.com}
}

\maketitle              
\begin{abstract}
Generative Design workflows have introduced alternative paradigms in the domain of computational design, allowing designers to generate large pools of valid solutions by defining a set of goals and constraints. However, analyzing and narrowing down the generated solution space, which usually consists of various high-dimensional properties, has been a major challenge in current generative workflows. By taking advantage of the interactive unbounded spatial exploration, and the visual immersion offered in virtual reality platforms, we propose V-Dream, a virtual reality generative analysis framework for exploring large-scale solution spaces. V-Dream proposes a hybrid search workflow in which a spatial stochastic search approach is combined with a recommender system allowing users to pick desired candidates and eliminate the undesired ones iteratively. In each cycle, V-Dream reorganizes the remaining options in clusters based on the defined features. Moreover, our framework allows users to inspect design solutions and evaluate their performance metrics in various hierarchical levels, assisting them in narrowing down the solution space through iterative cycles of search/select/re-clustering of the solutions in an immersive fashion. Finally, we present a prototype of our proposed framework, illustrating how users can navigate and narrow down desired solutions from a pool of over 16000 monitor stands generated by Autodesk's Dreamcatcher software.

\end{abstract}
\section{Introduction}
\begin{figure}
\centering
  \includegraphics[width=0.6\columnwidth]{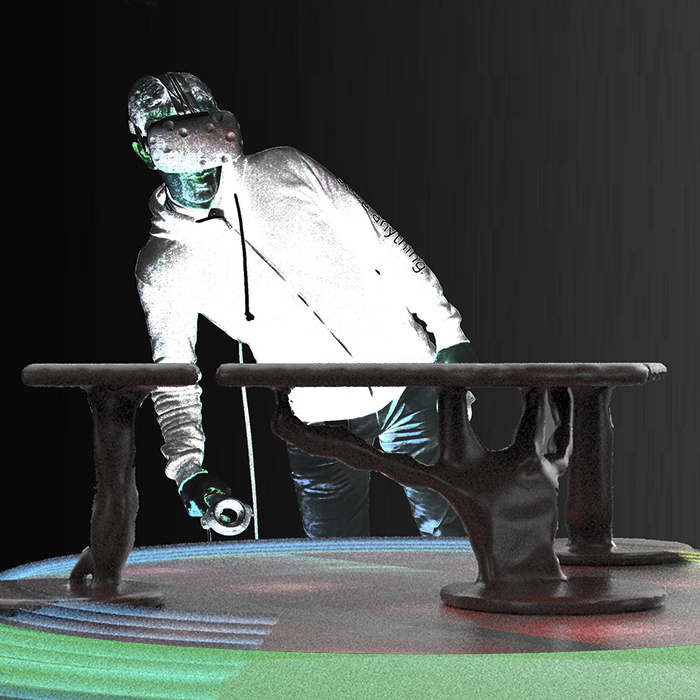}
  \caption{V-Dream allows generative designers to search, inspect, and re-organize the solution space within Virtual Reality.}~\label{fig:teaser}
\end{figure}

Advances in design automation and cloud-based computing has facilitated the rapid shift of computer-aided design paradigms towards generative workflows \cite{shea2005towards,mccormack2004generative}. Unlike traditional CAD-based processes, where a single design solution was modeled using a set of computational tools, in generative design workflows, designers specify high-level goals and constraints and, the system automatically generates large sets of solutions all corresponding to the defined design criteria. In addition to geometrical attributes, generative design systems can be integrated with performance evaluators \cite{shea2005towards,oxman2008performance} and simulation engines \cite{echenagucia2015early} to quantitatively assess and optimize the generated solution landscape.  With the availability of high-performance computing and cloud services, this process can be parallelized, allowing faster generation, improving the performance evaluation, and generating larger solution landscapes \cite{autodeskDream}. Users of such systems are then responsible to choose between plausible design candidates, which is often considered a complex task \cite{turrin2011design,von2012paragen}. These users should inspect the high-dimensional properties of each solution as well as assessing their aesthetic qualities.

 \begin{figure}
\centering
  \includegraphics[width=1\columnwidth]{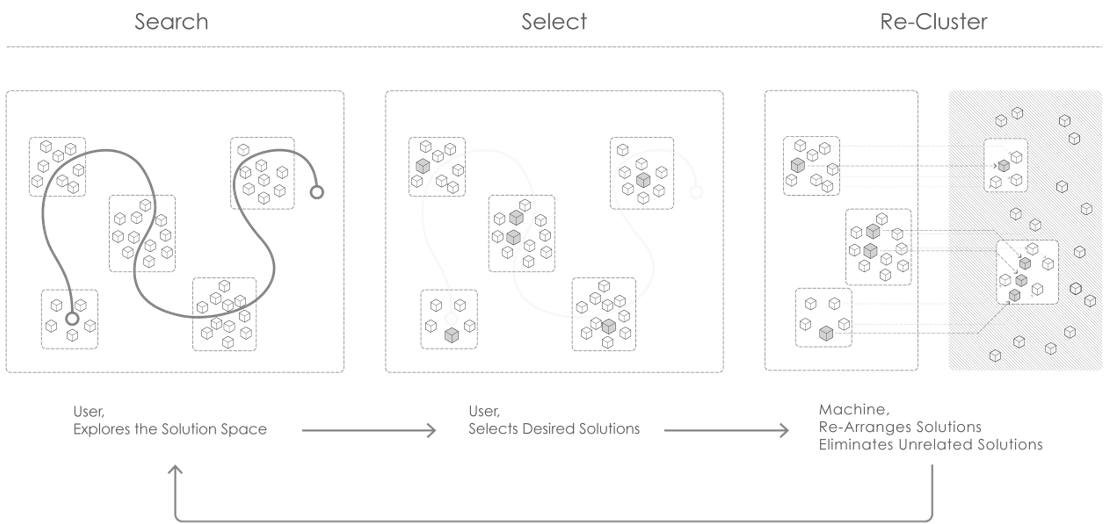}
  \caption{General workflow of the V-Dream framework: Performing exploratory search, labeling initial preferences of viable design solutions, re-clustering, and re-organizing the solution space based on user selection using a recommender system.}~\label{fig:workflow}
\end{figure}

There are two general approaches recognizable among the current generative design workflows: (i) Convergence generative design and (ii) divergence generative design \cite{matejka2018dream}. In convergence generative design, the search mechanism is implemented in a way to converge the solution space into a single solution or a set of limited solutions. However, depending on the level of clarity and accuracy of the goals, constraints, and the fitness function, the optimization process may dismiss many potentially acceptable solutions, which could have been otherwise chosen by the designer. Also, automated optimization methods do not leverage human expertise and can only find solutions that are optimal with regard to an invariably defined problem space \cite{scott2002investigating}. On the other hand, in divergent generative design, the whole solution space is generated, then the designers utilize sorting, clustering, and filtering tools to manually navigate and explore the solution space. Rather than looking at a limited set of solutions, designers have the chance to continually re-define their goals and constraints, allowing a more comprehensive control over the generative process. As divergent generative workflows often produce large numbers of solutions, organizing the solution space to effectively explore the data is considered a critical step in design explorations. 

The re-emergence of virtual reality (VR) technology as the next-generation human-computer interfaces has allowed various disciplines to explore how spatial interactions with virtual objects can benefit their field. VR can potentially elevate the sense of immersion and spatial awareness for users, providing designers the opportunity to interact with virtual objects in higher degrees of senses. Studies suggest that immersive environments enhance spatial understanding when compared to 2D or non-immersive 3D representations \cite{schnabel2003spatial,paes2017immersive}. Moreover, VR has been broadly used in various decision-making processes, from design tasks \cite{caldas2019design} to data analysis workflows \cite{chandler2015immersive}. Immersive content generated within virtual experience can support collaboration and allow users to visualize various mediums of data to assist real-world decision-making. 

In this paper, we introduce V-Dream, a virtual reality generative design exploration framework for analyzing large-scale solution spaces. Using V-Dream, users can navigate, organize, and cluster a solution space to locate or narrow down potential ideal design solutions from the generated outcomes. We use a hybrid approach of a user-centered stochastic search and a system-based recommender system to guide the user towards its desirable group of solutions. We believe the framework introduced in this paper can facilitate the integration of generative design workflows into next-generation spatial computing platforms. The exploration of large-scale data and the ability to intuitively perform spatial navigation tasks while analyzing design instances can be a promising step for the future generative design systems.

\begin{figure*}
\centering
  \includegraphics[width=1\columnwidth]{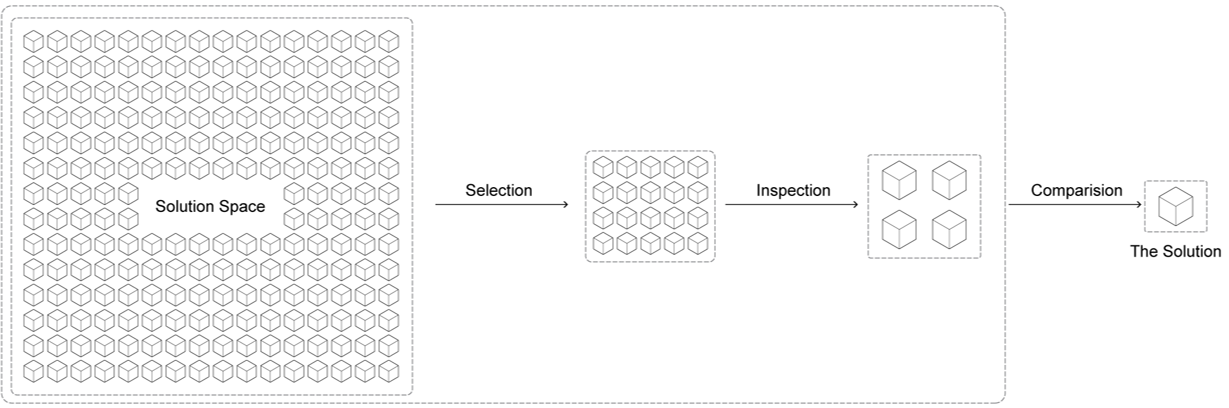}
  \caption{Generative design workflows allow designers to explore, select, inspect, and compare various examples from the solution space and narrow down to the desired one.}~\label{fig:process}
\end{figure*}

\section{Background}

\subsubsection{Dataset Exploration}
There is a rich body of literature allocated to the general task of multi-variate dataset exploration \cite{Koren2008,StuartMoore2006,hearst2008uis} for various applications. Such approaches have integrated filtering and recommender systems with data visualization methods to allows user to explore and search within a possible solution space. Interaction with large datasets has also been widely investigated and reviewed in user interface literature. Koyama et al. \cite{koyama2014crowd} present a unique approach to encoding the parameter space of an attribute directly on a slider widget, providing interactivity with the data bounds. The work of Kim et al. \cite{Kim2016} explores mapping multi-dimensional data onto a two-dimensional scatter plots. They address this by letting users drag examples to the extreme ends of the scatter plot to define how each axis should behave. We also incorporate a similar approach for bounding solutions spaces by allowing users to select desired data points, which would later reorganize the spatial visualization of the solution landscape.

\begin{figure}
\centering
  \includegraphics[width=0.95\columnwidth]{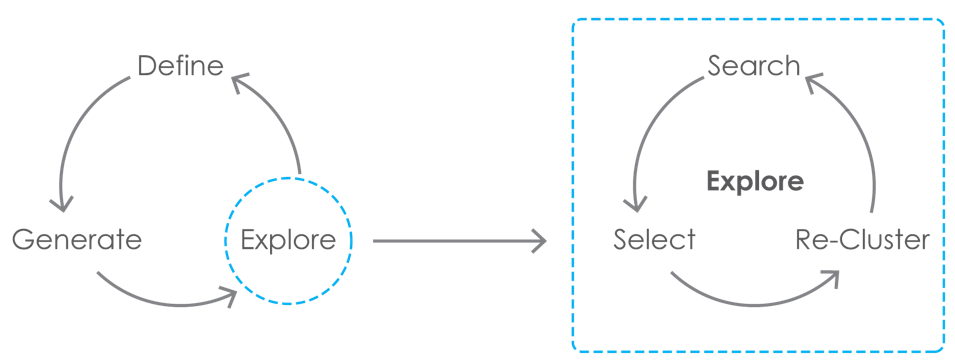}
  \caption{V-Dream allows designers to engage with an External Generation Loop, which iteratively defines, generates, and explores design solutions (left), while also interacting with an Internal Exploration Loop to filter, cluster, and re-organize the generated solution space(right).}~\label{fig:loop}
\end{figure}

\subsubsection{Generative Design Exploration}
Studies on how to interpret generative design solutions, and to provide the user with appropriate workflows to modify and interact with the generated solution space has been a topic of interest among various researchers \cite{scott2002investigating,natureHardy,meignan2015review}. Chaszar et al. \cite{Chaszar2016MultivariateIV} explored methods and tools for multivariate interactive data visualization of the generated designs and simulation results. They aimed at enabling designers to focus not only on high-performing results but also examine suboptimal ones. Mueller and Ochsendorf \cite{mueller2015combining}  proposed a computational design exploration approach that extends the existing interactive evolutionary algorithms to integrate the designers' preferences. They addressed this goal by allowing designers to set the evolutionary parameters, namely mutation rate, generation size, and parent selection. More recently, Bidgoli and Veloso \cite{bidgoli2019deepcloud} explored how generative systems could potentially learn both aspects of the problem definition and the design space through processing a database of existing solutions without the supervision of the designer. We use a similar approach in our clusterization steps while narrowing down the solution space based on users' previous selections of desired candidates.

\begin{figure}
\centering
  \includegraphics[width=0.7\columnwidth]{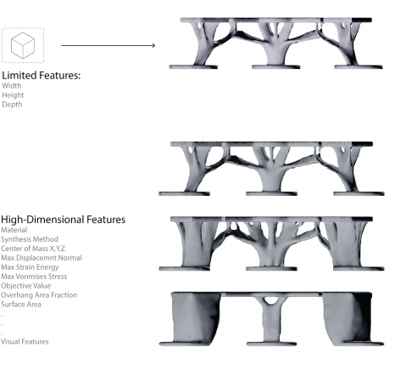}
  \caption{Limited feature analysis vs high-dimensional feature analysis.}~\label{fig:highDim}
\end{figure}

\subsubsection{Geometrical Interaction in VR}
Designing and modeling 3D geometry within VR interfaces provides the opportunity for designers to spatially interact with virtual models in real-life scale. Examples of such applications are Gravity Sketch VR \cite{gravitySketch}, Project Sugarhill \cite{sugarhill}, and vSpline \cite{arnowitz2017vspline}, which allow designers to implement conventional CAD modeling procedures within an immersive environment. Immersive painting and sculpting have also been widely explored in commercial applications such as Google Tilt Brush and Oculus Medium. Abbasi-Asl et al. \cite{abbasi2019brain} explored non-physical input modules by using a Brain-Computer Interface in VR for virtual spatial sculpturing. Innes et al. \cite{innes2017virtual} showed that designers tend to use significantly fewer actions in VR to achieve a similar result
than working on a monitor. Hsu et al.\cite{hsu2020design} developed a VR system that allows multiple design stakeholders to discuss about an architectural model in VR and modify the geometry during discussion using mid-air sketching as well as on-surface sketching in the virtual environment. Other examples of collaborative evaluation and modeling for architectural and design applications within VR can be seen in the work of \cite{greenwald2017multi,keshavarzi2019affordance,kunert2019multi}.

\subsubsection{Performance Evaluation in VR}
During the evaluation process, the user analyzes each solution concerning two distinct factors: 1) the aesthetic characteristics, 2) the simulated performance metrics. While the first factor is subjective and qualitative, the latter is quantitative and objective, which requires the user to utilize various methods to comprehend and assess effectively. In this regard, studies on immersive analytics \cite{chandler2015immersive,donalek2014immersive} have explored how VR platforms can be used to support scientific data visualization, analytical reasoning, and decision-making. For example, for building performance evaluation, Nytsch-Geusen et al. developed VR visualizations using bi-directional data exchange between energy simulation tools and the Unity game engine  \cite{Nytsch_2016}. The work of Rysanek et al. introduces a workflow for managing building information and performance data in VR with equirectangular image labeling methods \cite{Rysanek_Miller_Schlueter_2017}. Immersive interfaces have also been applied for structural investigations, finite element method simulations \cite{Hambli_Chamekh_Bel_Salah_2006}, and CFD visualizations \cite{berger2015cfd,malkawi2005new} and urban simulations \cite{cristie2015cityheat,sobral2019visualization}. Keshavarzi et al. demonstrated how interactive qualitative daylighting renderings can be overlaid with qualitative simulations within VR \cite{keshavarzi2019radvr}. We take advantage of such overlaying methods in our, where various performance metrics are visualized within the surrounding space of the target object.

\section{The V-Dream Framework}

\subsection{Design Goals}

The development of V-Dream is directed by two primary design goals. First, we aim to capture an iterative generative design workflow in which the designer can define, generate, and explore the design solution in multiple cycles. Each generative design cycle would allow the user to re-define the new cycle with adjusted goals and constraints. Second, we intend to provide a high dimensional exploration phase in which both qualitative and quantitative properties of individual solutions can be assessed. We elaborate more on each design goal bellow. 

\subsubsection{Iterative Generative Design Workflow}

Specifying the right design goals and constraints is a critical task in generative workflows. Failure to define a set of correct and effective design criteria would result in limited or inappropriate solution spaces, where ideal candidates may be excluded, and therefore, not evaluated by the user or the optimization module. However, depending on the complexity of the design and the expertise of the user, correctly defining such properties may
not happen in the early stages of design explorations. In such cases, users tend to redefine the objective functions to calibrate the solver to generate better solutions spaces. Therefore, we aim to design the generative design workflow in V-Dream as an iterative process, where generation, evaluation, and exploration are performed in a potentially cyclical workflow. Moreover, as generating each iteration of the solution space often requires heavy computation and time resources, minimizing the number of iterations using effective evaluation methods and recommender systems have also been explored during our development of V-Dream. In this regard, we aim to develop the exploration process as another internal cyclical workflow, where users can navigate, evaluate, and re-organize the generated solution space in each redefined generation. Figure \ref{fig:loop} illustrates the external generation loop and internal exploration loop of our proposed framework.

\subsubsection{Highlight Qualitative and Quantitative Properties}

During the conceptual design phase, designers must consider a wide range of performance-oriented goals. These include quantitative and measurable goals, such as structural efficiency, cost, and embodied energy, as well as qualitative goals that cannot be expressed numerically, such as aesthetics, constructability, and contextual appropriateness. 
The designer's early-stage responsibilities include balancing these requirements to attain a satisfying initial design concept. There is a wide range of tools and methods to support designers' decision-making during this phase when it comes to processing quantitative data. In contrast, designers have limited options while handling qualitative data and subjective measures. We aim to propose a means to address both aspects in the proposed framework. 

\subsection{Workflow}
In V-Dream, the search is an interactive process between the user and the algorithms in an immersive VR environment. The user relies on its subjective measures to initiate and direct the search process while the algorithm supports the process through a series of recommendations and reorganization of the solution space. Thus, a hybrid workflow that both relies on a stochastic approach and recommender system will be achieved which enhances the user's experience in searching within the solution space. 
The user starts from an exploratory search through the solution space, which is already clustered based on their visual and quantitative properties. Next, the user marks a limited number of solutions as the starting seeds. Then, the backend algorithm re-clusters and re-organizes the solution space based on the user's selection and eliminates irrelevant instances (Fig \ref{fig:workflow}). In the next sub-sections, we discuss the elements and steps of this workflow in more detail.
\subsection{Solution Clustering}\label{sec:clus}
Divergent generative workflows often produce significantly large pool of solutions. Accordingly, organizing the solution space to ensure a practical data exploration is a critical prerequisite for the design explorations phase. Therefore, we propose to utilize high-dimensional data clusterization methods, which have been widely explored in various data exploration applications. This approach facilitates organizing the solution space based on performance features that can be simulated or extracted as quantitative values. However, as discussed in the previous sections, qualitative aspects of each solution, such as aesthetics and form, also play an essential role in design decision-making. Such qualities commonly correspond to geometrical features of the solution space, and therefore clustering techniques integrated into the system should also reflect geometrical properties of the data. 

\begin{figure}
\centering
  \includegraphics[width=0.95\columnwidth]{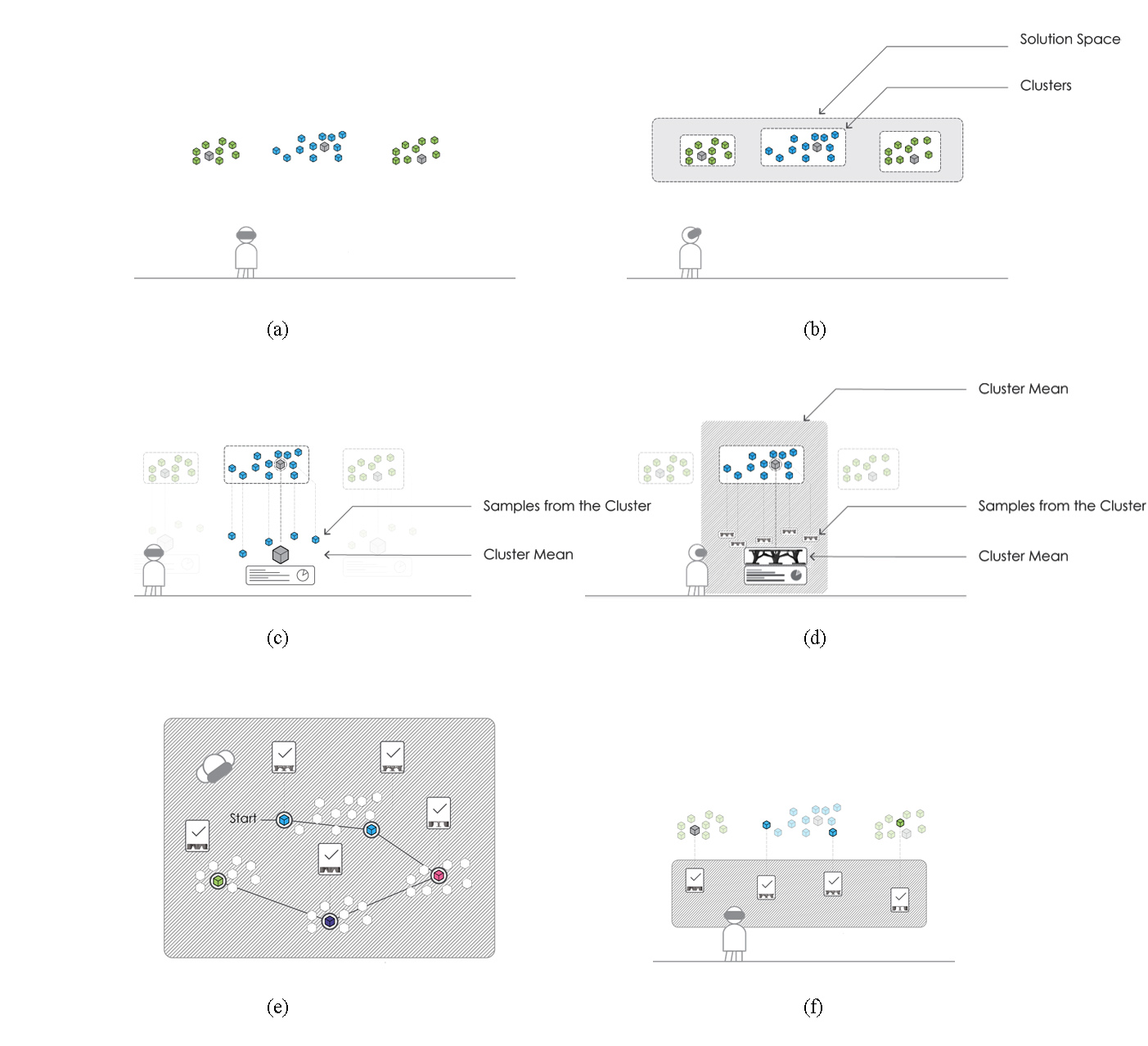}
  \caption{Six stages of our proposed hybrid search approach: a) initial solution generation and clusterization b) brief inspection of the solution space and clusters, c \& d) approaching a cluster and observing the mean member on the Visualization table while the other samples in the cluster  are visible as 3d objects e) exploring the space and adding various samples to its list of desired options, f)  triggering the re-clustering process to let the backend engine reorganize the clusters based on the user preferences.}~\label{fig:movement}
\end{figure}
We therefore propose to use shape clustering methods introduced by Huang et al. \cite{huang2012optimization}, which allows representative subsets to be extracted and optimized using a set of point-to-point correspondences between each of the representative shapes and the entire collection. Such an approach would allow users to explore the geometrical dimension of the solution spaces by initially evaluating cluster representatives of the generated solutions to maintain a general comprehension of the aesthetic properties of the corresponding subsets. If the user intends to explore similar solutions, the corresponding subset would expand and new clusters and representative shapes would be calculated and visualized within the bounds of the selected subset. This approach can iteratively execute until the user narrows down its geometrical search with the solution space.

To implement the clusterization exploration in VR space, as shown in Fig \ref{fig:movement}, we render the initial clustered results as a large room-scale map, above the user's height. The clusterization criteria are defined by the user and can be a combination of multiple performance factors or general geometrical properties, which organize the generated solution based on shape similarity. Each solution is rendered as a bright point, contributing to what is seen as a sky full of stars. Cluster representatives are positioned bellow corresponding subsets within the user's HMD height range and are visualized using the Visualization Tables described in the following section.

\subsection{Visualization Tables}\label{vizTables}
\begin{figure}
\centering
  \includegraphics[width=0.8\columnwidth]{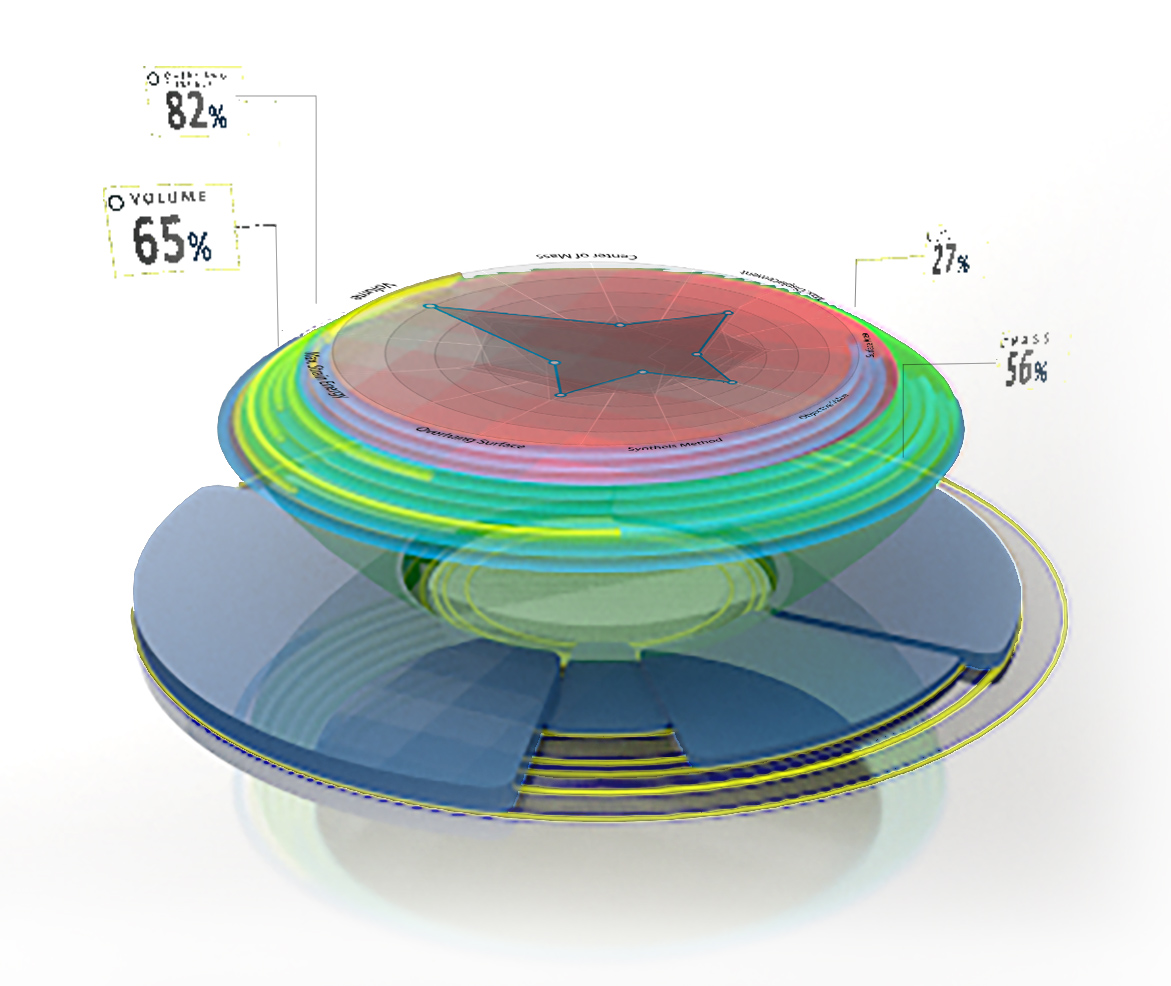}
  \caption{Visualization tables used for quantitative assessment of generated solutions.}~\label{fig:table}
\end{figure}

To allow the user to perform quantitative analyses of the generated solutions, V-Dream places each 3D model on a \textit{Visualization Table}. As shown in Fig \ref{fig:table}, Visualization Tables plot a spider diagram on the top side of the surface, while generating radial graphs on the perimeter. Such property would allow the user to compare various quantitative factors while inspecting the aesthetic and visual quality of the object itself, which is placed on top of the table. A Visualization tables can be rotated in it's place allowing the user to assess the generated object the data visualization without needing to move around the table virtually.

\subsection{Navigation Strategy}
The clustering algorithm maps the high-dimensional solution space into a 3-dimensional space. This allows clusters to be represented as a cloud of solutions in an immersive environment, where similar solutions are distributed in close proximity of each other to form a cluster. Meanwhile, larger distances between clusters imply the significant variance between them. 
Users can navigate and explore the solution space, visit clusters, and inspect each instance separately. However, the sheer size of the solution space renders it impossible to visit and inspect each instance independently. Finding a feasible yet inclusive method to efficiently navigate the solution space is a critical task that needed to be addressed. We discuss two different approaches for solution space navigation: 1) navigation based on stochastic approach, and 2) recommender system assisted navigation. Finally, we propose a hybrid approach that combines the two to form an optimal workflow.

\subsubsection{Stochastic Navigation}
In stochastic navigation, the user starts from an arbitrary point in the solution space and navigates through it, inspecting one solution at a time. The organization of solutions in clusters based on their visual similarity allows the user to move from one cluster to its neighbors and gradually find the desired solution. During our initial user interviews, we found that this approach is popular among designers and users that are willing to explore and examine a wider range of options while developing a comprehensive mental map of solution space. This mental map provides these users with a  sense of orientation and location for further navigation of solution space.

\begin{figure*}
\centering
  \includegraphics[width=1\columnwidth]{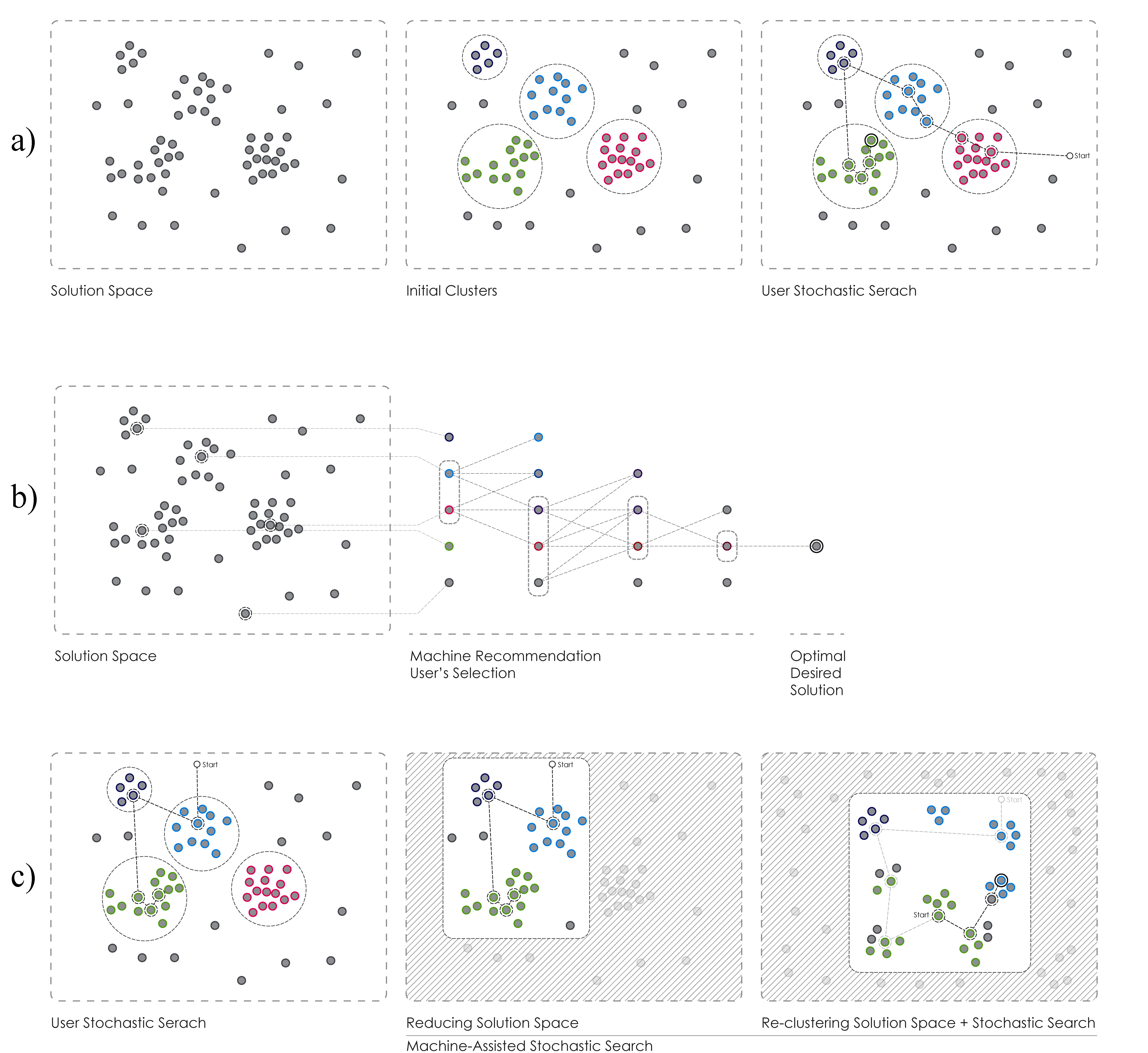}
  \caption{Abstract illustration of the search process a) stochastic search b) recommender system c) hybrid approach.}~\label{fig:conceptFig}
\end{figure*}

\begin{figure}
\centering
  \includegraphics[width=0.87\columnwidth]{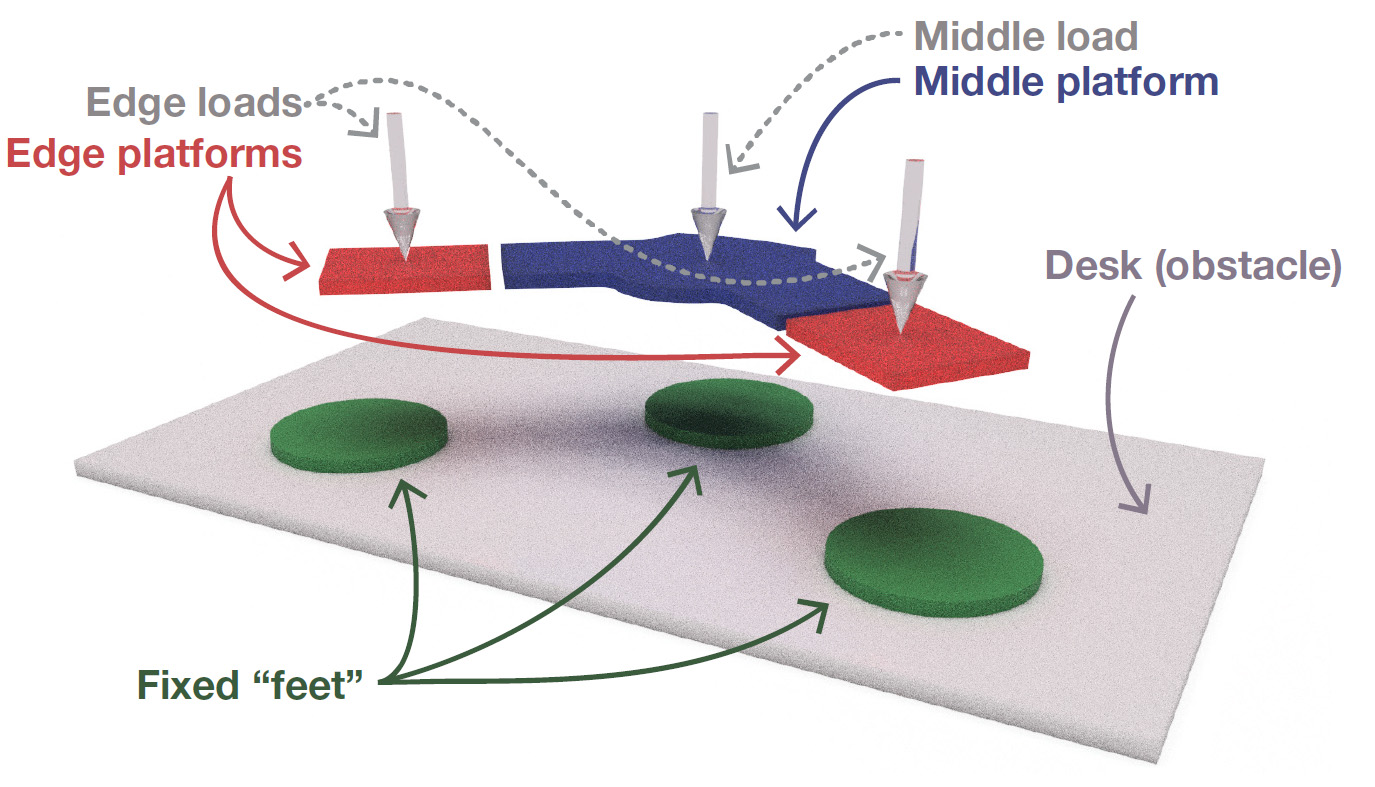}
  \caption{Problem definition describing the locations of the feet, platforms, and desk surface geometry, and the position and direction of the static forces. (Image from Matejka et al. \cite{matejka2018dream})}~\label{fig:monitorStand}
\end{figure}

\subsubsection{Recommender System Navigation}
In contrast with the stochastic approach, navigation based on a recommender system aims to utilize recommender algorithms to help the user efficiently narrowing down the solution space. In this approach, the algorithm provides the user with the first round of seeds, representing the most populated clusters of the solutions space to select from. After the first round, the algorithm searches for clusters with similar properties to the ones that the user has already selected from. The algorithm also eliminates non-relevant clusters from the pool. It then recommends the user with another round of seeds. This iteration continues until the user is satisfied with one or a limited selection of options from the solution space.

\subsubsection{Hybrid Navigation}
While both approaches have their owns pros and cons, a comprehensive and optimal search approach might be a hybrid of these two. We propose a hybrid approach that starts with a stochastic search,  followed by a recommender algorithm. This approach potentially helps the user to develop a mental image of solution space while collecting the first round of seeds. Furthermore, the user can activate the recommender algorithm to 1) eliminate unrelated solutions, 2) find new seeds from the most relevant ones to the current seeds, and 3) re-cluster the rest of the solutions space based on the fresh seeds. 
In this approach, users determine their desired features to measure the similarities between the instances of solution space. These features can be visual characteristics, performance data, or any high-dimensional metadata that comes from generative design backend. Machine learning methods can be used to extract visual features, while analytical methods can be used for quantitative and qualitative ones. Using dimension reduction methods (i.e., T-SNE \cite{maaten2008visualizing}, PCA \cite{jolliffe1986principal}) features can be mapped on a 3D space while various clustering methods (discussed in Sec \ref{sec:clus}) can determine the cluster boundaries. The user repeats the Search-Select-Re-cluster cycle until it finds the desired solution.

\subsection{Hierarchical Interface}
\begin{figure*}
\centering
  \includegraphics[width=1\columnwidth]{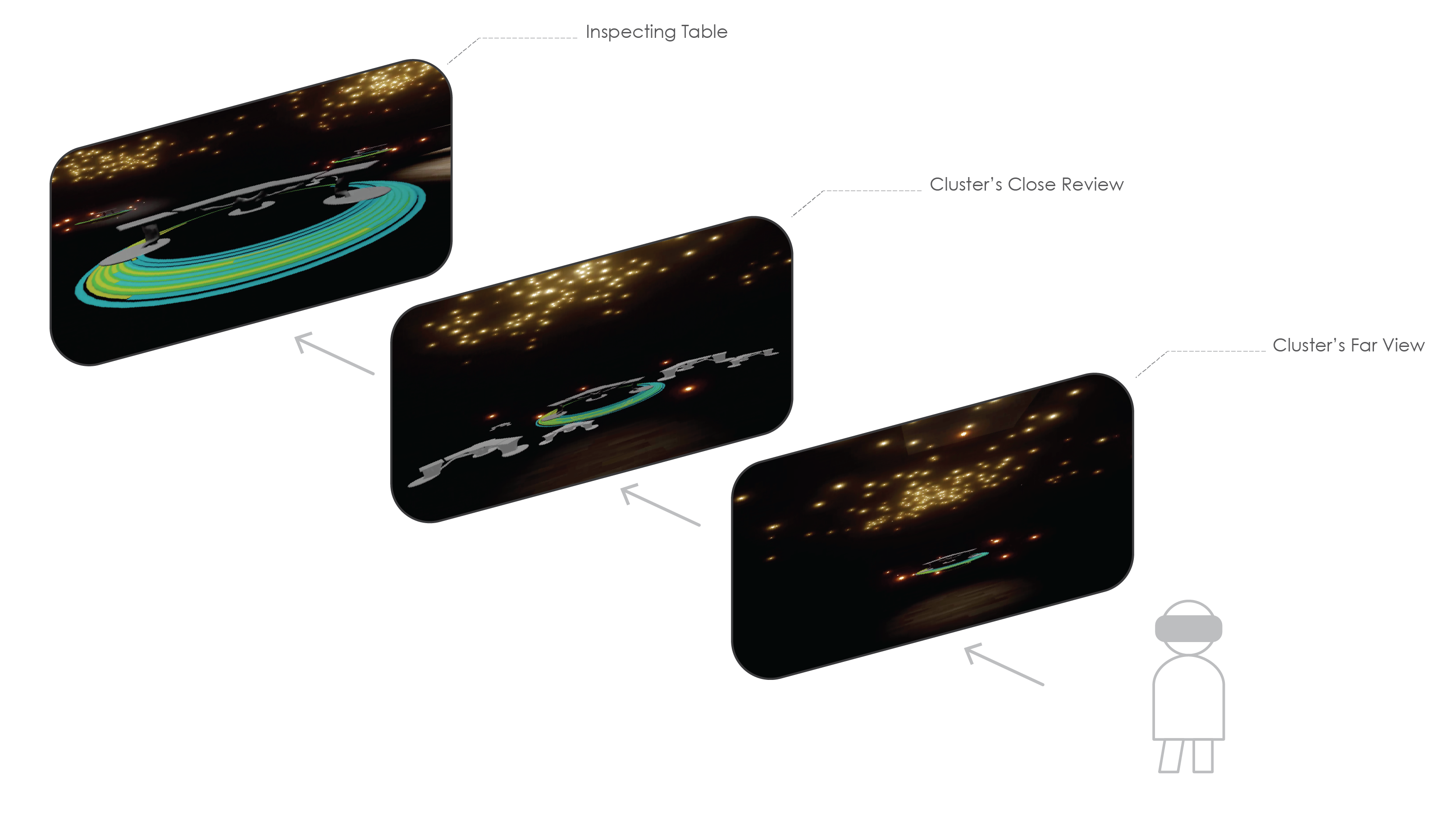}
  \caption{Hierarchical visualization process: the three resolution stages for solution exploration and assessment.}~\label{fig:heirach}
\end{figure*}

Given the large number of generated solutions that surround the user during spatial navigation, we design a hierarchical interface in which objects and their visualization tables are rendered with different levels of details in relation to their distance with the user. Clusters are initially represented by a single solution located on the Visualization table and surrounded by other solutions of the cluster rendered as bright points. Once the user navigates towards the cluster, the minimized solution points would expand to individual Visualization tables holding their corresponding objects. As the user navigates closer to each table, additional metrics and detailed values of the quantitative data would be visible, allowing detailed evaluation of the generated solutions. Such a hierarchical approach would allow the user to explore large datasets in various visualization scales, providing a chance to spatially organize the generated solutions for its stochastic assessment. Fig \ref{fig:heirach} illustrates three levels of the hierarchical process explained above within our developed prototype. 

\subsection{Re-clustering}

After the user selects a number of desired solutions, the re-clustering of the solution spaces using the recommender system is executed based on the aesthetic properties of the selected solutions. With such an approach, we integrate qualitative factors of the generated solutions with the generative design cycle, providing an additional design goal of the generative design definition. After each search/select/re-cluster cycle, users are more likely to view desirable solutions in their surrounding navigation space, which align with selections they had made in the previous cycles. In other words, the re-organization of the generated clusters would allow users to thoughtfully navigate the generated solution space, and decide whether an additional re-definition of the whole design space is necessary.

\section{Prototype}
To demonstrate V-Dream in action, we developed a Virtual Reality prototype to explore a pre-generated solution space generated using Autodesk Dreamcatcher \cite{autodeskDream}. Dreamcatcher is an internally developed experimental generative design platform for engineering design problems where multiple shapes and topology optimization algorithms are employed to synthesize model geometries that optimally satisfy defined criteria \cite{allaire2004structural,bendsoe2013topology}. V-Dream is built on Autodesk's Stingray game engine. To provide custom modifications by potential users, we developed the framework on Stingray's Visual Programming Language (VPL) tool while programming the main functionalities using Lua, Stingray's primary programming API language. It is important to note that the Autodesk's Stingray software was discontinued after the initial development of this prototype. We evaluate our developed system by interviewing six user participants who are involved in Generative Design software design and development at Autodesk Research. We use an HTC Vive Pro for the VR experience.

\subsection{Dataset}

The design exploration dataset we used was initially introduced by \cite{matejka2018dream}, formulating a design problem that was (i) easy to understand without any specialized domain knowledge, (ii) could be used by a broad range of people, and (iii) was a problem that could be satisfied by a wide variety of designs options. The design task corresponding to the dataset generation was to create a 3D printed monitor stand and to raise the monitor 80mm off the surface of a desk. The geometry consists of three disks at the bottom, which represent where the monitor stand will contact the desk, and three flat sections placed 80mm above the desk, which trace the shape of the monitor’s base. The static weight of the monitor is defined 8.3kg, so the stand should be able to support a load of at least 200N, and the ability to support heavier loads is desirable as well.

The weight of the monitor is modeled as two independent forces: a middle load on the middle platform and an outer load distributed evenly to the two outside platforms. In addition to the geometry of the design, the following properties are calculated: Center of Mass (x, y, z), Weight, Overhang Percentage, Surface Area, Area/Volume Ratio, as well as properties related to the performance under the load conditions: Maximum Displacement, Max. Strain, Total Strain, Max. Vonmises, and Objective Value (for the simulation). Four parameters in the problem definition and solver configuration are varied in this dataset as follows: the middle load, the outer load, voxel size, and volume minimization. Combining these factors leads to 16,800 total designs. The generated solutions contain the 3D geometry of each design, a metadata file describing all the input parameters used, and properties calculated for each design.

\section{Discussions}

During our user interviews and prototyping testing, we observe the stochastic approach is popular among designers that are willing to explore and examine a broader range of options while developing a larger mental image of solution space. In contrast, using the recommender system to reduce the exploration time is preferable among users that value performance more than aesthetics, where qualitative properties play a less important role in their design decision-making. While our proposed framework intends to capture both exploration approaches, the fact that the recommender system plays a primary role in re-clustering the solution space to guide the user to its preferable solution is not always convincing for our surveyed users. The recommender system, in theory, is highly dependent on both the design problem and user input, and therefore, cannot guarantee the ideal re-clustering of the solution space. Such limitation may cause unwanted disruptions in the design exploration of the generative designer, or misguide the user towards non-desirable results.  

In some instances, we found participants to navigate through the solution space and update their selection seed with new choices. Same as a shopping experience, we observed them delete the previous selections seed and replace them with new design options. This emphasized the need for a thorough exploration of generated solutions, in which some better solutions can be found once correct navigation is found.  Furthermore, as expected, we observed users to simply walk by (virtually navigate) from some cluster representatives in which they found not as much attractive, or not as desirable than previously selected solutions, and in contrast, stopped and surveyed clusters in which they expected to find desirable solutions. 

While V-Dream primarily targets generative design workflows, we believe the proposed framework can be a promising module for other search applications that require large-scale data exploration in future spatial computing interfaces. Applications such as immersive shopping where users intend to explore different product types and categories while assessing and analyzing the pros and cons of each individual product can be seen as an example of such integration. Social media platforms can also be integrated with hybrid exploration mechanisms proposed in this study, allowing thorough navigation between user-profiles and generated content in a targeted fashion.

\section{Conclusion and Future Work}

In this paper, we introduce an immersive exploration framework for analyzing large-scale generative design solution spaces. By targeting divergent generative design workflows, we propose a user-centric stochastic navigation approach coupled with a system-based recommender system to guide the user towards its desirable set of solutions. V-Dream allows users to assess quantitative and qualitative properties of generated solutions by organizing, clustering, and navigating the generated solution space. We believe the framework proposed in this paper can enhance design search and exploration by utilizing the spatial benefits of next-generation spatial computing platforms. Targeted exploration of large scale data integrated with machine-guided workflows introduced in our proposed framework can be a promising step for future generative design systems. 

Our work, of course, comes with limitations. First, our developed prototype currently does not allow generative design goals and constraints to be defined directly in the virtual reality interface. Generative design solutions are generated and transferred from Dreamcatcher in an offline manner. Maintaining a bi-directional data-workflow between other generative design software (Fusion 360, Revit, Dreamcatcher) can be considered future work of our approach. Furthermore, running comprehensive user studies to evaluate how generative design users interact and design with our proposed workflow is a necessary next step. In addition, given the discontinuity of Autodesk's Stingray engine in which our prototype was developed, we hope to transfer our prototype to another platform (Unity or Unreal) in order to share our codebase with the community.

%
%
%
\bibliographystyle{splncs04}
%

\bibliography{template}




\end{document}